\documentclass[runningheads]{llncs}
\usepackage{epsfig}
\usepackage{graphicx}
\usepackage{amsmath,amssymb}
\usepackage{color}
\usepackage{latex_pkgs/maths}
\usepackage{latex_pkgs/msformatting}
\usepackage[width=122mm,left=12mm,paperwidth=146mm,height=193mm,top=12mm,paperheight=217mm]{geometry}
\begin{document}

\pagestyle{headings}
\mainmatter
\title{TraMNet - Transition Matrix Network for Efficient Action Tube Proposals} 
% Replace with your title

\titlerunning{TraMNet - Towards Flexible Action Tube  Proposals}
% Replace with a meaningful short version of your title
%
\author{Gurkirt Singh \and
Suman Saha \and Fabio Cuzzolin}
%
%Please write out author names in full in the paper, i.e. full given and family names. 
%If any authors have names that can be parsed into FirstName LastName in multiple ways, please include the correct parsing, in a comment to the volume editors:
%\index{Lastnames, Firstnames}
%(Do not uncomment it, because you may introduce extra index items if you do that, we will use scripts for introducing index entries...)
\authorrunning{Gurkirt Singh, Suman Saha and Fabio Cuzzolin}
% Replace with shorter version of the author list. If there are more authors than fits a line, please use A. Author et al.
%

\institute{Oxford Brookes University, UK\\
\email{gurkirt.singh-2015@brookes.ac.uk}}
\maketitle % typeset the header of the contribution
%%%%%%%%%%%%%%%%%%%%%%%%%%%%%%%%%%%%%%%%%%%%%%
\begin{abstract}  
Current state-of-the-art methods solve spatio-temporal action localisation by extending 2D anchors to 3D-cuboid proposals on stacks of frames, to generate sets of temporally connected bounding boxes called \textit{action micro-tubes}. However, they fail to consider that the underlying anchor proposal hypotheses should also move (transition) from frame to frame, as the actor or the camera do. Assuming we evaluate $n$ 2D anchors in each frame, then the number of possible transitions from each 2D anchor to he next, for a sequence of $f$ consecutive frames, is in the order of $O(n^f)$, expensive even for small values of $f$.

To avoid this problem we introduce a \textbf{Tra}nsition-\textbf{M}atrix-based \textbf{Net}work (TraMNet) which relies on computing transition probabilities between anchor proposals while maximising their overlap with ground truth bounding boxes across frames, and enforcing sparsity via a transition threshold. As the resulting transition matrix is sparse and stochastic, this reduces the proposal hypothesis search space from $O(n^f)$ to the cardinality of the thresholded matrix. At training time, transitions are specific to cell locations of the feature maps, so that a sparse (efficient) transition matrix is used to train the network. At test time, a denser transition matrix can be obtained either by decreasing the threshold or by adding to it all the relative transitions originating from any cell location, allowing the network to handle transitions in the test data that might not have been present in the training data, and making detection translation-invariant. Finally, we show that our network is able to handle sparse annotations such as those available in the DALY dataset, while allowing for both dense (accurate) or sparse (efficient) evaluation within a single model. We report extensive experiments on the DALY, UCF101-24 and Transformed-UCF101-24 datasets to support our claims.
\end{abstract}
\section{Introduction} \label{sec:intro}

Current state-of-the-art spatiotemporal action localisation works~\cite{saha2017amtnet,kalogeiton2017action,hou2017tube} focus on learning a
spatiotemporal multi-frame 3D representation by extending frame-level 2D object/action detection approaches ~\cite{Georgia-2015a,Weinzaepfel-2015,girshick-2014,ren2015faster,liu15ssd,Saha2016,peng2016eccv,singh2016online}. 
These networks learn a feature representation from pairs~\cite{saha2017amtnet} or chunks~\cite{kalogeiton2017action,hou2017tube} 
of video frames, allowing them to implicitly learn the temporal correspondence between inter-frame action regions (bounding boxes).
As a result, they can predict micro-tubes~\cite{saha2017amtnet} or tubelets~\cite{kalogeiton2017action}, 
i.e., temporally linked frame-level detections for short subsequences of a test video clip.
Finally, these micro-tubes are linked~\cite{saha2017amtnet,kalogeiton2017action,hou2017tube} 
in time to locate action tube instances~\cite{singh2016online} spanning the whole video.% sequence.  

\begin{figure}[t!]
  \centering
  %\vspace{-0.2cm}
  \includegraphics[scale=0.19]{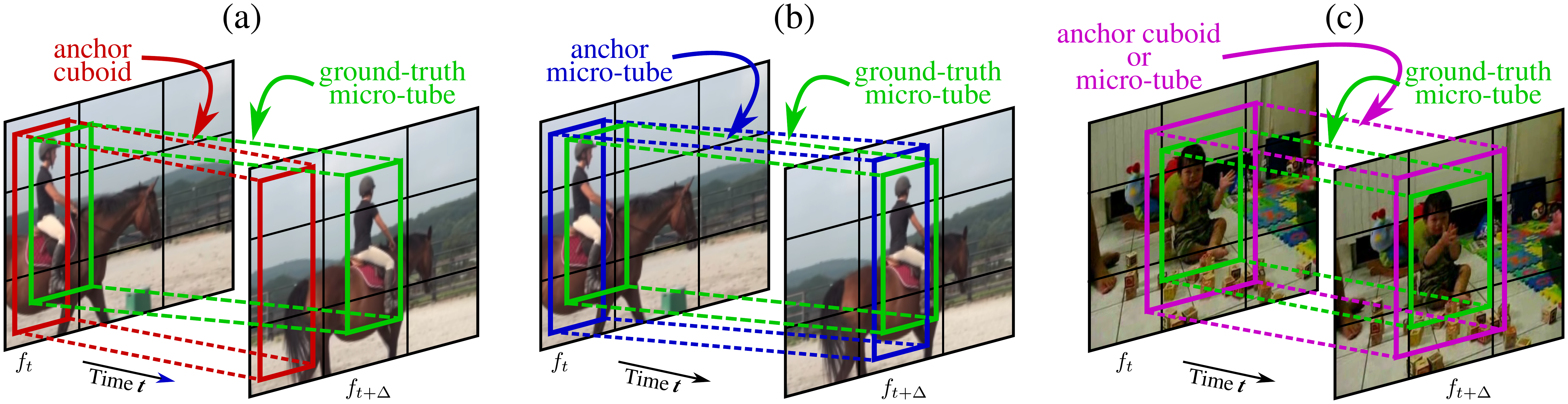}
  \vspace{-0.2cm}
  \caption{
    {\small
      \textit{
      Illustrating the key limitation of anchor cuboids using a ``dynamic'' action like
      ``horse riding''. \textbf{(a)} A horse rider changes its location from frame $f_t$ to $f_{t+\Delta}$
      as shown by the ground truth bounding boxes (in green).
      As the anchor cuboid generation~\cite{saha2017amtnet,kalogeiton2017action} is constrained by the spatial location of the 
      anchor box in the first frame $f_t$, the overall spatiotemporal IoU overlap 
      between the ground-truth micro-tube and the anchor cuboid is relatively low.
      \textbf{(b)} In contrast, our anchor micro-tube proposal generator is much more flexible,
      as it efficiently explores the video search space via an approximate
      transition matrix estimated based on a hidden Markov model (HMM) formulation. As a result,
      the anchor micro-tube proposal (in blue) generated by the proposed model
      exhibits higher overlap with the ground-truth.
      \textbf{(c)} For ``static'' actions (such as ``clap'') in which
      the actor does not change location over time, 
      anchor cuboid and anchor micro-tubes have the same % Fabio: the same or just similar? Gurkirt: same
      spatiotemporal bounds.
            }
    }
 }
\label{fig:opt_prop} \vspace{-0.7cm}
\end{figure}

%---------------------------------------------------------------------
% two main concersn with the current soa
% 1) extend anchor box to anchor cuboid which is not 
% the way to exploit the video search space in an optimal way
% we proose to use transition matrix for that
% 2) do not support sparse annotation
%---------------------------------------------------------------------
These approaches, however, raise two major concerns.
Firstly, 
they~\cite{saha2017amtnet,kalogeiton2017action,hou2017tube} 
generate action proposals by extending
2D object proposals (anchor/prior boxes for images)~\cite{liu15ssd,ren2015faster} 
to 3D proposals (anchor cuboids for multiple frames) (cf. Fig.~\ref{fig:opt_prop}~\textbf{(a)}).
This cannot, by design, provide an optimal set of training 
hypotheses, as the video proposal search space ($\mathcal{O}(n^f)$) is much larger than the image proposal search space ($\mathcal{O}(n)$), 
%By extending 2D anchor boxes to 3D cuboids, we still limiting the proposal generation to 
%image search space i.e. of $\mathcal{O}(n)$, whereas the actual action proposal search space for a video is of $\mathcal{O}(n^f)$, % n x f x k^f
where $n$ is the number of anchor boxes per frame and
$f$ is the number of video frames considered.
% $k$ is the tube smoothness factor i.e.
% the number of spatial neighbours an anchor box wish to link in the next frame.
%The video search space grows exponentially as the number of video frames $f$ increases.
% effectively exploit the video search space by generating 
Furthermore, 
%unlike object detection where 2D anchor boxes show promising results~\cite{ren2015faster,liu15ssd}, the 
3D anchor cuboids are very limiting for action detection purposes.
Whereas they can be suitable for ``static'' actions (e.g. ``handshake'' or ``clap'', 
in which the spatial location of the actor(s) does not vary over time),
they are most inappropriate for ``dynamic'' ones (e.g. ``horse riding'', ``skiing'').
Fig.~\ref{fig:opt_prop} %illustrates the key limitation of the anchor cuboids.
underscores this issue.
%Note that for a ``dynamic'' action like 
For ``horse riding'', for instance, allowing ``flexible'' anchor micro-tubes (as those generated by our approach, 
Fig.~\ref{fig:opt_prop}~\textbf{(b)}) much improves
the spatio-temporal overlap %between the anchor micro-tube (generated by our approach) and 
with the ground-truth 
(Fig.~\ref{fig:opt_prop}~\textbf{(a)}).
\emph{Designing a deep network which can effectively make use of the video search space to
generate high-quality action proposals, while keeping the computing cost as low as possible}, is then highly desirable.
{
To this end, we produced a new action detection dataset which is a ``transformed'' version of UCF-101-24~\cite{soomro-2012},
in which we force action instances to be dynamic (i.e., to change their spatial location significantly over time) by introducing random translations in the 2d spatial domain.
We show that our proposed action detection approach outperforms the baseline~\cite{saha2017amtnet} when trained and tested on this transformed dataset.
}

In the second place,
action detection methods such as~\cite{kalogeiton2017action,hou2017tube} 
require dense ground-truth annotation for network training:
bounding-box annotation is required for $k$ consecutive video frames,
where $k$ is the number of frames in a training example.
Kalogeiton~\etal~\cite{kalogeiton2017action} use $k=6$ whereas 
for Hou~\etal~\cite{hou2017tube} $k=8$.
Generating such dense bounding box annotation for long video sequences is 
highly expensive and impractical ~\cite{daly2016weinzaepfel,ava2017gu}.
The latest generation action detection benchmarks DALY~\cite{daly2016weinzaepfel} 
and AVA~\cite{ava2017gu}, in contrast, provide sparse bounding-box annotations.
More specifically,  DALY has $1$ to $5$ frames bounding box annotation per action instance irrespective of the duration of an instance, 
whereas AVA has only one frame annotation per second.
\emph{This motivates the design of a deep network able to handle sparse annotations, 
while still being able to predict micro-tubes over multiple frames}.

Unlike~\cite{kalogeiton2017action,hou2017tube}, % where $k$ consecutive video frames are used  as training examples,
Saha~\etal~\cite{saha2017amtnet} recently proposed to use \emph{pairs} of successive frames
$(f_t, f_{t+\Delta})$, eliminating the need for dense training annotation when 
$\Delta$ is large e.g. $\Delta=\{ 5,10,21\}$ or arbitrary DALY~\cite{daly2016weinzaepfel}.
If the spatio-temporal IoU (Intersection over Union) overlap between the ground-truth micro-tube 
and the action proposal could be improved (cf. Fig.~\ref{fig:opt_prop}),
such a network would be able to handle sparse annotation
(e.g., pairs of frames which are $\Delta=21$ apart). Indeed, the use of 
pairs of successive frames $(f_t, f_{t+\Delta})$ in combination with the flexible anchor proposals introduced here, is arguably more efficient than any other state-of-the-art method~\cite{saha2017amtnet,kay2017kinetics,hou2017tube} for handling sparse annotations (e.g. DALY~\cite{daly2016weinzaepfel} and AVA~\cite{ava2017gu}).
.
%----------this can be removed if space is limited -----
%This idea of improving the overlap between anchor and ground-truth boxes is intuitive.
%Redmon~\etal~\cite{redmon2016yolo9000} already demonstrated better object detection performances
%by improving the overlap of anchor boxes by k-means.
%------------------------------
%% WE NEED TO SHORTEN THIS
\begin{figure*}[t]
  \centering
  % \vspace{-0.2cm}
  \includegraphics[scale=0.2]{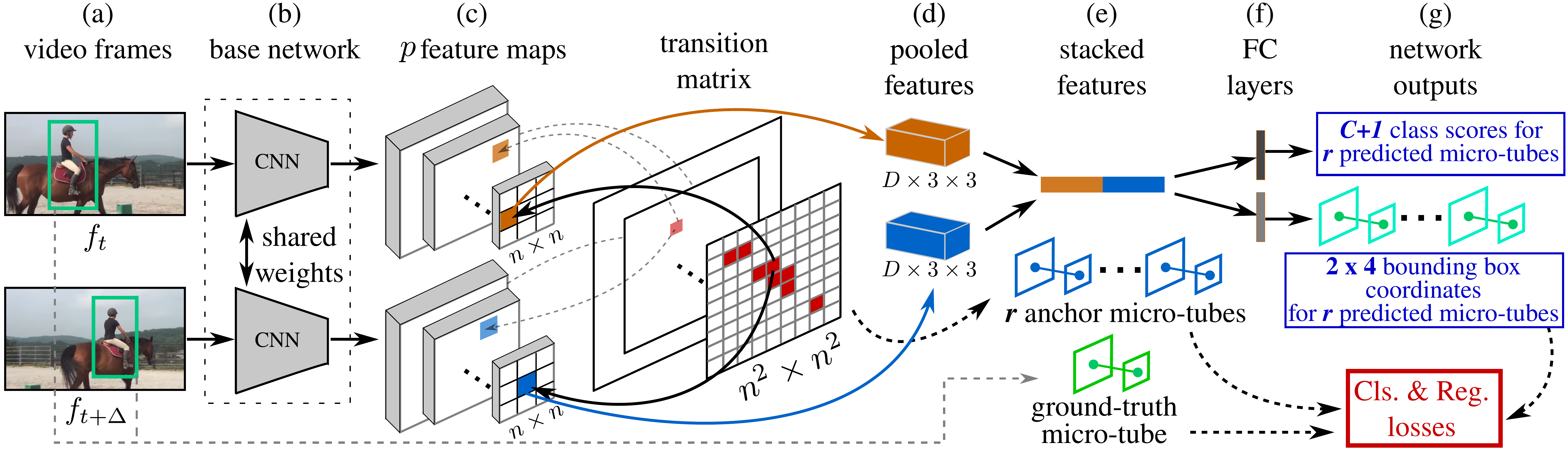}
 \vspace{-0.4cm}
  \caption{
    {\small
      \textit{Overview of our proposed TraMNet at training time. The diagram is described in the text.}
    }
 }
\label{fig:algorithmOverview} 
\vspace{-0.7cm}
\end{figure*}
\\
\textit{\textbf{Concept.}} 
Here we support the idea of constructing training examples using pairs of successive frames. %~\cite{saha2017amtnet}.
However, the model we propose is able to generate a rich set of action proposals 
(which we call \emph{anchor micro-tubes}, cf. Fig.~\ref{fig:opt_prop}) 
using a transition matrix (cf. Section~\ref{subsec:transition_matrix})
estimated from the available training set. 
Such transition matrix encodes the probability of a temporal link between
an anchor box at time $t$ and one at $t+\Delta$,
and is estimated within the framework of discrete state/continuous observation
hidden Markov models (HMMs, cf. Section~\ref{subsec:hmm})~\cite{elliott2008hidden}. 
Here, the hidden states are the 
2D bounding-box coordinates $[x_{min}, y_{min}, x_{max}, y_{max}]'$
of each anchor box from a (finite) hierarchy of fixed grids at different scales.
The (continuous) observations are the kindred four-vectors of coordinates associated with 
the ground truth bounding boxes 
(which are instead allowed to be placed anywhere in the image).
%------------OVERVIEW------
%Unlike ~\cite{saha2017amtnet,kalogeiton2017action,hou2017tube},
%our proposed approach generates action proposals 
%(which we call \emph{anchor micro-tubes}, cf. Fig.~\ref{fig:opt_prop})
Anchor micro-tubes are not bound to be strictly of cuboidal (as in~\cite{saha2017amtnet,kalogeiton2017action,hou2017tube}) shape, 
thus giving higher IoU overlap with the ground-truth, specifically
for instances where the spatial location of the actor changes significantly 
from $f_t$ to $ f_{t+\Delta}$ in a training pair.
We thus propose a novel configurable deep neural network architecture
(see Fig.~\ref{fig:algorithmOverview} and Section~\ref{sec:methodology})
which leverages high-quality micro-tubes shaped by learnt anchor transition probabilities.
%generated from the inter-frame transition probabilities between 2D anchor boxes.
% Fabio: we already said this
%As the transition-matrix is computed in a supervised way from the training data, this drastically reduces the video search space from $\mathcal{O}(n^f)$ to the cardinality of transition-matrix $|A|$.

% The proposed model shows promising results even when trained on sparse annotations and evaluated in a dense manner.
We quantitatively demonstrate that the resulting action detection framework: 
\textbf{(i)} is suitable for datasets with temporally sparse frame-level bounding box annotation (e.g. DALY~\cite{daly2016weinzaepfel} and AVA~\cite{ava2017gu});
\textbf{(ii)} outperforms the current state-of-the-art~\cite{saha2017amtnet,kalogeiton2017action,singh2016online} by exploiting the anchor transition probabilities learnt from the training data. 
\textbf{(iii)} {is suitable for detecting highly `dynamic' actions (Fig.~\ref{fig:opt_prop}), as shown by its outperforming the baseline~\cite{saha2017amtnet} when trained and tested on the ``transformed'' UCF-101-24 dataset.}
%Finally, we will demonstrate how transition matrix helps when transition we more obvious with the help of 
\iffalse
We show the impact of transition matrix estimation on the 
``transformed'' UCF-24 dataset, which we generated by applying random translations to existing annotations and images for a number of classes, to artificially simulate random action instance shift across frames.
\fi
\\
\textbf{Overview of the approach.}
Our network architecture %adapts 
builds on some of the architectural components of~\cite{liu15ssd,saha2017amtnet,kalogeiton2017action} (Fig.~\ref{fig:algorithmOverview}).
The proposed network takes as input a pair of successive video frames $f_t, f_{t+\Delta}$ 
(where $\Delta$ is the inter-frame distance) (Fig.~\ref{fig:algorithmOverview}~\textbf{(a)})
and propagates these frames through 
a base network comprised of two parallel CNN networks
(\S~\ref{subsec:base_network} Fig.~\ref{fig:algorithmOverview}~\textbf{(b)}),
which produce two sets of $p$ conv feature maps
$K^{t}_{p}$ and $K^{t+\Delta}_{p}$ forming a pyramid.
These feature pyramids are used by a configurable pooling layer 
(\S~\ref{subsec:reconfig_pooling_layer} and Fig.~\ref{fig:algorithmOverview}~\textbf{(d)})
to pool features based on the transition probabilities defined by a 
transition matrix $\mathbf{A}$ 
(\S~\ref{subsec:transition_matrix}, Fig.~\ref{fig:algorithmOverview}).
The pooled conv features are then stacked 
(\S~\ref{subsec:reconfig_pooling_layer} and Fig.~\ref{fig:algorithmOverview}~\textbf{(e)}),
and
the resulting feature vector is passed to two parallel fully connected (linear) layers
(one for classification and another for micro-tube regression, see \S~\ref{subsec:conv_2_linear} and Fig.~\ref{fig:algorithmOverview}~\textbf{(f)}),
which predict the output micro-tube 
and its classification scores for each class $C$~\textbf{(g)}.
Each training mini-batch is used to compute the classification and
micro-tube regression losses given 
the output predictions, 
ground truth and anchor micro-tubes.
We call our network \emph{``configurable''} because 
the configuration of the pooling layer (see Fig.~\ref{fig:algorithmOverview}~\textbf{(d)})
depends on the transition matrix $\mathbf{A}$, 
and can be changed by altering the threshold applied to $\mathbf{A}$ 
(cf. Section~\ref{subsec:transition_matrix}).
or by replacing the transition matrix with a new one for another dataset.
\\
\textbf{Contributions.}
%\red{[Note for Gurkirt: please write %the contrib.]}\\
In summary, 
we present a novel deep learning architecture for spatio-temporal action localisation 
which:
\begin{itemize}
     \vspace{-0.2cm}
     \item introduces an efficient and flexible anchor micro-tube hypothesis generation framework to generate high-quality action proposals;
     \item handles significant spatial movement in dynamic actors without penalising more static actions;
     \item is a scalable solution for training models on both sparse or dense annotations.
     %\item we introduce transformed UCF-24 dataset with random translation applied to annotations and images, to simulate artificial translation in actor's position across two frames. 
\end{itemize}
\vspace{-0.6cm}

\section{Related work}
Traditionally, spatio-temporal action localisation was widely studied using local or figure centric features~\cite{vanGemert2015apt,oneata2014efficient,jain2014tublet,sapienza2014,Sultani2016}. Inspired by Oneata~\etal~\cite{oneata2014efficient} and Jain~\etal~\cite{jain2014tublet}, Gemert~\etal~\cite{vanGemert2015apt} used unsupervised clustering to generate 3D tubelets using unsupervised frame level proposals and dense trajectories. 
As their method is based on dense-trajectory features~\cite{wang-2011}, however, it fails to detect actions characterised by small motions~\cite{vanGemert2015apt}.

Recently, inspired by the record-breaking performance of CNNs based 
object detectors~\cite{redmon2016yolo9000,ren2015faster,liu15ssd} several scholars
~\cite{singh2016online,Saha2016,Georgia-2015a,peng2016eccv,Weinzaepfel-2015,weinzaepfel2016towards,zolfaghari2017chained} 
tried to extend object detectors to videos for spatio-temporal action localisation.
These approaches, however, fail to tackle spatial and temporal reasoning jointly at the network level, 
as spatial detection and temporal association are treated as two disjoint problems.
Interestingly, Yang \etal \cite{yang2017spatio} use features from current, frame $t$ proposals to `anticipate' region proposal locations in $t+\Delta$ and use them to generate detections at time $t+\Delta$, thus failing to take full advantage of the anticipation trick to help with the linking process.
\\ More recent works try to address this problem by predicting micro-tubes~\cite{saha2017amtnet} or tubelets~\cite{kalogeiton2017action,hou2017tube} for a small set of frames taken together. As mentioned, however, these approaches use anchor hypotheses which are simply extensions of the hypothesis in the first frame, thus failing to model significant location transitions. In opposition, 
here we address this issue by proposing anchor regions which move across frames, 
%depending upon transition of max overlapping anchor proposal from one frame to other. 
as a function of a transition matrix estimated at training time from anchor proposals of maximal overlap.

Advances in action recognition are always going to be helpful in action detection from a general representation learning point of view. For instance,
Gu \etal \cite{ava2017gu} improve on \cite{peng2016eccv,kalogeiton2017action} by plugging in the inflated 3D network proposed by \cite{carreira2017quo} as a base network on multiple frames.
Although they use a very strong base network pre-trained on the large ``kinetics''~\cite{kay2017kinetics} dataset, they do not handle the linking process within the network as the AVA~\cite{ava2017gu} dataset's annotations are not temporally linked.  
%fail to handle sparse annotation of AVA dataset to improve the linking process using their dense network.

Temporal association is usually performed by some form of ``tracking-by-detection'' ~\cite{singh2016online,Weinzaepfel-2015,Georgia-2015a} of frame level detections.
Kalogeiton \etal~ \cite{kalogeiton2017action} adapts the linking process proposed by Singh \etal~\cite{singh2016online} to link tubelets, whereas Saha \etal~\cite{saha2017amtnet} builds on \cite{Georgia-2015a} to link micro-tubes. 
Temporal trimming is handled separately either by sliding window \cite{daly2016weinzaepfel,peng2016eccv}, or in a label smoothing formulation solved using dynamic programming~\cite{Saha2016,Evangel-2014}. 
For this taks we adopt the micro-tube linking from~\cite{kalogeiton2017action,singh2016online} and the online temporal trimming from \cite{singh2016online}. 
We demonstrate that the temporal trimming aspect does not help on UCF101-24 (in fact, it damages performance), while it helps on the DALY dataset in which only 4\% of the video duration is covered by action instances.

% Fabio: what is this paragraph for? Does not look very relevant
%Object detection in videos has also been getting popular \cite{kang2016tcnn,feichtenhofer2017detect}, 
%after ImageNet's video object detection challenge (VID) \cite{russakovsky2015imagenet}.
%Kang \etal \cite{kang2017tpn} use multiple frames with cuboid anchor boxes to generate tubelets and pool features from tubelet regions and pass them to an LSTM for classification. 
%Feichtenhofer \etal \cite{feichtenhofer2017detect} use the frame proposals at time $t$ to generate the expected position of detections frame $t+k$ and use that for linking and detection at $t+k$. Ground truth at $t+k$ is used for anticipation at time $t$ as a target, hence such a method requires dense annotations for consecutive frames.
%forthis still could provide the sub-optimal solution and requires dense annotations to train. % Fabio: this one needs fixing too
%Last but not least, video object detection does not require temporal trimming or association, as the final evaluation in VID is per frame.
% deep recofigurable feature pooling for action detection
\section{Methodology}~\label{sec:methodology}
\begin{figure*}[t]
  \centering
  \includegraphics[scale=0.19]{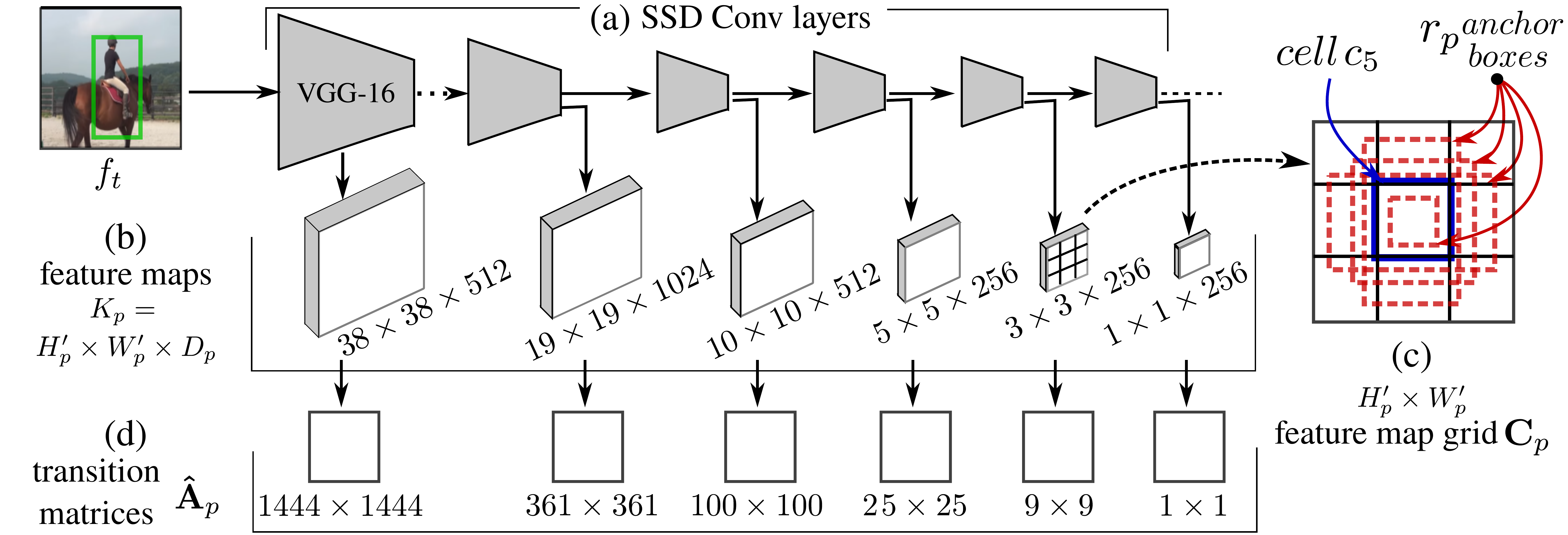}
  \vspace{-0.2cm}
  \caption{
    {\small
      \textit{Base network architecture.
      \textbf{(a)} SSD convolutional layers;
      \textbf{(b)} the corresponding conv feature maps outputted by each conv layer;
      \textbf{(c)} $r$ anchor boxes with different aspect ratios
      assigned to cell location $c_5$ of the $3 \times 3$ feature map grid;
      \textbf{(d)} transition matrices for the $P$ feature map grids in the pyramid, where $P=6$.
      }
    }
 }
\label{fig:base_net}
\vspace{-0.7cm}
\end{figure*}
%----------------------------------------------------------------------------
% \vspace{-0.2cm}
In Section~\ref{subsec:base_network},
we introduce the base network architecture used for feature learning.
We cast the action proposal generation problem
in a hidden Markov model (HMM) formulation (\S~Section~\ref{subsec:hmm}),
and introduce an approximate estimation
of the HMM transition probability matrix using
a heuristic approach (\S~Section~\ref{subsec:transition_matrix}).
% ( Section~\ref{subsec:hmm}),
The proposed approximation is relatively inexpensive
and works gracefully (\S~\ref{subsec:discussion}). % TODO refer experiments with trans matrix
%----------------------------------------------------------------------------
In Section~\ref{subsec:reconfig_pooling_layer},
a configurable pooling layer architecture is presented
which pools convolutional features from the regions in the two frames linked by the estimated transition probabilities.
%----------------------------------------------------------------------------
Finally, the output layers of the network (i.e., the micro-tube
regression and classification layers) are described in Section~\ref{subsec:conv_2_linear}.

% in the base network tell that this is SSD based,
% draw a diagram for the add conv layers and the 6 feature maps
% then show the anchor for only one cell location in the 5x5 grid

\subsection{Base network} \label{subsec:base_network}
The base network takes as inputs a pair of video frames
$(f_t,f_{t+\Delta})$ and propagates them through two parallel CNN streams
(cf. Fig.~\ref{fig:algorithmOverview}~\textbf{(b)}).
In Fig.~\ref{fig:base_net}~\textbf{(a)},
we show the network diagram of one of the CNN streams; the other follows the same design.
\\
The network architecture is based on Single-Shot-Detector (SSD)~\cite{liu15ssd}.
The CNN stream outputs a set of $P$ convolutional feature maps
$K_{p}$, $p = \{ 1,2, ..., P = 6 \}$
(feature pyramid, cfr. Fig.~\ref{fig:base_net}~\textbf{(b)})
of shape $[H'_{p} \times W'_{p} \times D_p]$,
where $H'_{p}$, $W'_{p}$ and $D_p$ are the height, width and depth
of the feature map at network depth $p$, respectively.
%Indeed the feature maps are extracted at different depths $p = \{ 1,2, ..., 6 \}$ of the network.
For $P=6$ %different network depth levels,
the conv feature map spatial dimensions are $H' = W' = \{ 38, 19, 10, 5, 3, 1\}$, respectively.
The feature maps at the lower depth levels (i.e., $p=1,2$ or 3)
are responsible for encoding smaller objects/actions,
whereas feature maps at higher depth levels encode
larger actions/objects.
For each cell location $c_{ij}$ of $[H'_{p} \times W'_{p}]$ feature map grid $\mathbf{C_p}$,
$r$ anchor boxes (with different aspect ratios)
are assigned where $r_{p} = \{ 4, 6, 6, 6, 4, 4 \}$.
E.g.
% at network depth $p=5$,
% we extract a conv feature map of shpae $[3 \times 3 \times 256]$,
at each cell location of the $3 \times 3$ grid in the pyramid,
$4$ anchor boxes are produced
(Fig.~\ref{fig:base_net}~\textbf{(c)}),
resulting in a total of $3 \times 3  \times 4 = 36$ anchor boxes.
These anchor boxes, assigned for all $P=6$ distinct feature map grids,
are then used to generate action proposal hypotheses based on the
transition probability matrix, as explained below.

Note that the proposed framework is not limited to any particular base network architecture,
and is flexible enough to accommodate any latest network~\cite{lin2017focal,carreira2017quo}. 
% with greater representation power.

\subsection{HMM-based action proposal generation} \label{subsec:hmm}
A \emph{hidden Markov model} (HMM) models a time series of (directly measurable) \emph{observations} $\mathbf{O} = \{ \mathbf{o}_1, \mathbf{o}_2, ..., \mathbf{o}_T\}$, either discrete or continuous, as randomly generated at each time instant $t$ by a \emph{hidden state} $\mathbf{q}_t \in \mathbf{Q} = \{ \mathbf{q}_1, \mathbf{q}_2, ..., \mathbf{q}_{N} \}$, whose series form a Markov chain, i.e., the conditional probability of the state at time $t$ given $\mathbf{q}_1,...,\mathbf{q}_{t-1}$ only depends on the value of the state $\mathbf{q}_{t-1}$ at time $t-1$. The whole information on the time series' dynamics is thus contained in a \emph{transition probability matrix}
$\mathbf{A} = [p_{ij}; i,j=1,..,n]$,
where $p_{ij} = P(\mathbf{q}_j|\mathbf{q}_i)$ is the probability of moving from state $i$ to state $j$, and $\sum^{N}_{j=1} p_{ij} = 1$ $\forall i$.
\\
In our setting, a state $\mathbf{q}_n$ is a vector containing the 2D bounding-box coordinates
of one of the anchor boxes
$[x^{a}_{min}, y^{a}_{min}, x^{a}_{max}, y^{a}_{max}]'$ %Fabio: why the superscript ``a''?
in one of the grids forming the pyramid (\S~\ref{subsec:base_network}).
The transition matrix encodes the probabilities of a temporal link existing between
an anchor box (indexed by $i$) at time $t$ and another anchor box (indexed by $j$) at time $t+\Delta$. The continuous observations $\mathbf{o}_t$, $t=1,...,T$ are the ground-truth bounding boxes, so that $\mathbf{O}$ corresponds to a ground-truth action tube.

In hidden Markov models, observations are assumed to be Gaussian distributed given a state $\mathbf{q}_i$, with mean $\mathbf{o}_\mu^i$ and covariance $\mathbf{Q}_\Sigma^i$.
After assuming an appropriate distribution for the initial state, e.g. $P(\mathbf{q}_0)\sim \mathcal{N}(0,I)$, the transition model $A = [P(\mathbf{q}_j|\mathbf{q}_i)]$ allows us to predict at each time $t$ the probability $P(\mathbf{q}_t | \mathbf{O}_{1:t})$ of the current state given the history of previous observations, i.e.,
the probability of each anchor box at time $t$ given the observed (partial)
ground-truth action tube.
Given a training set, the optimal HMM parameters ($A$, $\mathbf{o}_\mu^i$ and $\mathbf{Q}_\Sigma^i$ for $i=1,...,N$) can be learned using standard expectation maximisation (EM) or the Baum-Welch algorithm, by optimising the likelihood of the predictions $P(\mathbf{q}_t | \mathbf{O}_{1:t})$ produced by the model.
\\
%TODO: Fabio please look at these three paragraphs see if we save space - OK
Once training is done, at test time,
the mean
$\mathbf{o}_\mu^{\hat{\mathbf{q}}_t}$ of the conditional distribution of the observations given the state associated with the predicted state $\hat{\mathbf{q}}_t \doteq \arg \max_i P(\mathbf{q}_i|\mathbf{O}_{1:t})$ at time $t$
%of the learnt state distribution $\mathbf{Q}_{\mu}$
can be used to initialise the anchor boxes for each of the $P$ CNN feature map grids (\S~\ref{subsec:base_network}).
% for a training pair $(f_t,f_{t+\Delta})$
The learnt transition matrix $\mathbf{A}$ can be used to
generate a set of training action proposals hypotheses (i.e., anchor micro-tubes, Fig.~\ref{fig:opt_prop}).
As in our case the mean vectors $\mathbf{o}_\mu^i$, $i=1,..., N$ are known a-priori
(as the coordinates of the anchor boxes are predefined
for each feature map grid, \S~\ref{subsec:base_network}),
%we initialise the mean of the hidden state  $\mathbf{Q}_{\mu}$ with the anchor box coordinates
we do not allow the M-step of EM algorithm to update $\mathbf{Q}_{\mu} = [\mathbf{o}_\mu^i, i=1,...,N]$. Only the covariance matrix $\mathbf{Q}_{\Sigma}$ is updated.
% update the $Q_{\mu}$ at m-step of EM algorithm,

\subsection{\textbf{Approximation of the HMM transition matrix}}  \label{subsec:transition_matrix}

Although the above setting perfectly formalises the anchor box-ground truth detection relation over the time series of training frames,
a number of computational issues arise. At training time,
some states (anchor boxes)
may not be associated with any of the observations (ground-truth boxes) in the E-step, leading to
zero covariance for those states. % resulting in a non-positive definite covariance matrix.
%In addition, we observe during the HMM training that
Furthermore, for a large number of states (in our case $N = 8732$ anchor boxes),
it takes around $4$ days to complete a single HMM training iteration.
\\
%In response, we propose an approximation of HMM learning based on a heuristic approach.
In response, we propose to approximate the HMM's transition probability matrix $\mathbf{A}$ with
a matrix $\mathbf{\hat{A}}$ generated by a heuristic approach explained below.

\begin{figure*}[h]
  \vspace{-0.4cm}
  \centering
  \includegraphics[scale=0.16]{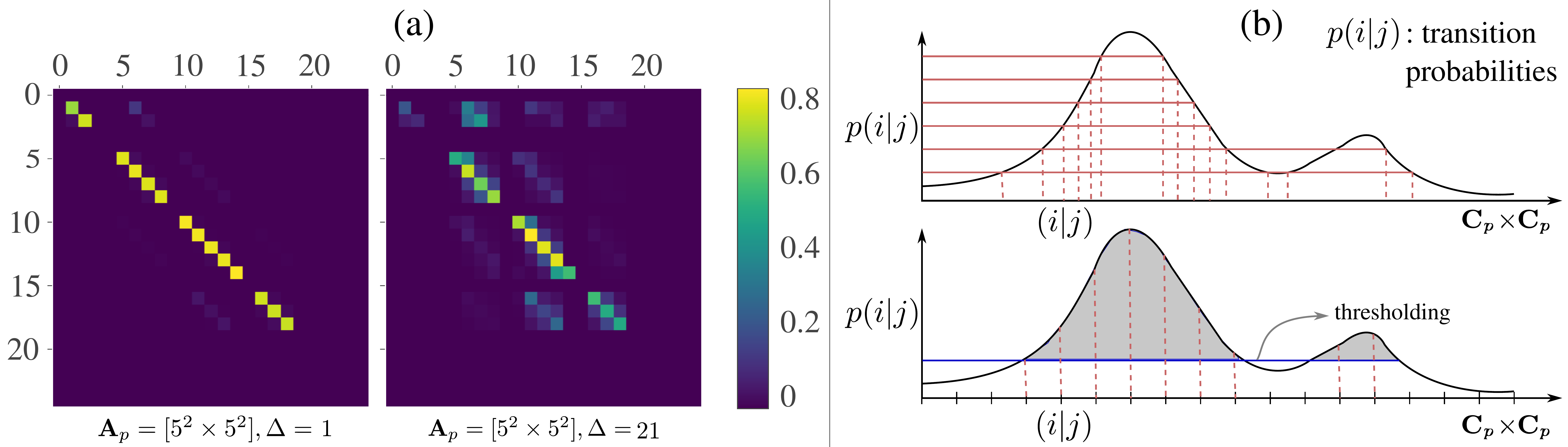}
   \vskip -0.4cm
  \caption{
    {\small
      \textit{
      \textbf{(a)}
      Transition matrix for a $5 \times 5$ feature map grid ($p=4$)
      for different $\Delta$ values.
      As $\Delta$ increases,
      off-diagonal probability values also increase,
      indicating a need for anchor micro-tubes rather than anchor-cuboids.
      \textbf{(b)} {Top - Monte Carlo sampling of transition hypotheses $(i,j) \in \mathbf{C}_{p} \times \mathbf{C}_{p}$ based on uniformly sampling the $[0,1]$ range. Bottom - our anchor micro-tube sampling scheme, based on thresholding the transition probabilities $p(i|j)$, is also stochastic in nature and emulates Monte Carlo sampling. The blue line denotes the threshold and the shaded area above the threshold line shows the sampling region, a subset of the product grid $\mathbf{C}_{p} \times \mathbf{C}_{p}$.}
      }
    }
 }
\label{fig:trans_mat}
\vspace{-7mm}
\end{figure*}

The problem is to learn a transition probability, i.e.,
the probability of a temporal link (edge)
between two anchor boxes $\{b^{a}_{t}, b^{a}_{t+\Delta}\}$
belonging to two feature map grids
$\mathbf{C}^{t}_{p}$ and $\mathbf{C}^{t+\Delta}_{p'}$.
%There are $p$ such pairs of grids corresponding to $p$ feature maps outputted by the base network (\S~\ref{subsec:base_network} Fig.~\ref{fig:base_net}~\textbf{(b)}).
If we assume that transitions only take place between states at the same level $p = p'$ of the feature pyramid, the two sets of anchor boxes
$\mathcal{B}^{t}_{p} = \{ b^{a}_{t_{1}}, ..., b^{a}_{t_{N}} \}$ and
$\mathcal{B}^{t+\Delta}_{p} = \{b^{a}_{{(t+\Delta)}_{1}}, ..., b^{a}_{{(t+\Delta)}_{N}}\}$
belonging to a pair of grids
$\{\mathbf{C}^{t}_{p},\mathbf{C}^{t+\Delta}_{p}\}$
are identical, namely:
$\mathcal{B}^{t}_{p} = \mathcal{B}^{t+\Delta}_{p} \doteq \mathcal{B}_{p} = \{b^a_i, i=1,...,N\}$,
allowing us to remove the time superscript. % $t$ and $t+\Delta$ and represent the set of all anchor boxes as $\mathcal{B}_{p}$ for clarity.
Recall that each feature map grid $C_{p}$ has spatial dimension $[H'_{p} \times W'_{p}]$.
\\
We compute a transition probability matrix $\mathbf{\hat{A}}_{p}$ \emph{individually for each grid level} $p$, resulting in $p$ such matrices of shape $[(H'_{p})^2 \times (W'_{p})^2]$ (see Fig.~\ref{fig:base_net}~\textbf{(d)}).
For example, at level $p=5$ we have a $3 \times 3$ feature map grids, so that
the transition matrix $\mathbf{\hat{A}}_{p}$ will be $[3^2 \times 3^2]$.
Each cell in the grid is assigned to $r_{p}$ anchor boxes, resulting in
$n = H'_{p} \times W'_{p} \times r_{p}$ total anchor boxes per grid
(\S~\ref{subsec:base_network}).
%Note that here we are interested in learning a transition probability between
%two anchor boxes %$b^{a}_{loc_{1}}$ and $b^{a}_{loc_{2}}$
%belonging to two different cell locations
%%%%%$c_{loc_{1}}$, $c_{loc_{2}}$
%in the grid, % where $loc_{1} \neq loc_{2}$ thus,
%thus further reducing the search space complexity from
%$\mathcal{O}(n^f)$ to $\mathcal{O}(L^f)$
%where, $L$ is the number of cell locations in a grid.
\vspace{-0.25cm}
\paragraph{\textbf{Transition matrix computation.}}
Initially, all entries of the transition matrix are set to zero:
$\mathbf{\hat{A}}[i,j] = 0$.
Given a ground-truth micro-tube $\mathbf{m}^{g} = \{b^{g}_{t}, b^{g}_{t+\Delta}\}$
(a pair of temporally linked ground-truth boxes~\cite{saha2017amtnet}),
%belonging to the training set,
we compute the IoU overlap for each ground-truth box
with all the anchor boxes $\mathcal{B}_{p}$ in the considered grid, namely:
$IoU(b^{g}_{t}, \mathcal{B}_{p})$ and $IoU(b^{g}_{t+\Delta}, \mathcal{B}_{p})$.
We select the pair of anchor boxes
%$\mathbf{m}^{a} = \{ b^{a}_{loc_{1}}, b^{a}_{loc_{2}} \}$
$\mathbf{m}^{a} = \{ b^{a}_{i}, b^{a}_{j} \}$
(which we term \emph{anchor micro-tube})
having the maximum IoU overlap with $\mathbf{m}^{g}$,
where $i$ and $j$ are two cell locations.
If $i = j$ (the resulting anchor boxes are in the same location) we get an anchor cuboid, otherwise a general anchor micro-tube.
\\
%%and we only consider pairs if $loc_{1} \neq loc_{2}$ i.e. $i \neq j$.
This is repeated for all $P$ feature map grids $\mathbf{C}_{p}$
to select the anchor micro-tube $\mathbf{m}^{a}_{p}$ with the highest overlap.
The best match anchor micro-tube $\mathbf{m}^{a}_{\hat{p}}$
for a given ground-truth micro-tube $\mathbf{m}^{g}$
is selected among those $P$, and
%and we vote for the corresponding cells (anchor regions) $i$ and $j$ by updating 
the transition matrix is updated as follows:
%$\mathbf{\hat{A}}[loc_{1}, loc_{2}] = \mathbf{\hat{A}}[loc_{1}, loc_{2}] + 1$.
$\mathbf{\hat{A}}[i, j] = \mathbf{\hat{A}}[i, j] + 1$.
The above steps are repeated for all the ground-truth micro-tubes in a training set.
Finally, each row of the transition matrix $\mathbf{\hat{A}}$ is normalised by dividing
each entry by the sum of that row.
\\
Fig.~\ref{fig:trans_mat} plots the transition matrix $\mathbf{\hat{A}}_p$ for $p=4$ (a feature map grid $5 \times 5$), for different values of $\Delta$.
As explained in the following, the configurable pooling layer
employs these matrices to pool conv features for action proposal classification and regression.

Although our approach learns transition probabilities for anchor boxes
belonging to the same feature map grid $\mathbf{C}_{p}$, we realise that
the quality of the resulting action proposals could be further
improved by learning transitions between anchors
across different levels of the pyramid. %$\mathbf{C}_{p}$s. 
As the feature dimension of each map varies in SSD, e.g. 1024 for $p=2$ and 512 for $p=1$, a more consistent network such as FPN~\cite{lin2017focal} with Resnet~\cite{he2016deep} would be a better choice as base architecture. Here we stick to SSD to produce a fair comparison with~\cite{kalogeiton2017action,singh2016online,saha2017amtnet}, and
leave this extension to future work.

\subsection{Configurable pooling layer} \label{subsec:reconfig_pooling_layer}
The SSD~\cite{liu15ssd} network uses convolutional kernels of dimension
$[3 \times 3 \times D]$ as classification and regression layers (called \emph{classification} and \emph{regression heads}).
More specifically, SSD uses $r \times 4$ kernels for bounding box regression
(recall $r$ anchor boxes with different aspect ratios are assigned to each cell location (\S~\ref{subsec:base_network}))
and $(C+1) \times r$ kernels for classification over the $p$ conv feature maps
(\S~\ref{subsec:base_network}). This is fine when the number of proposal hypotheses
is fixed (e.g., for object detection in images, the number of anchor boxes is set to $8732$).
In our setting, however,
the number of proposals varies depending upon the cardinality of transition matrix
$| \mathbf{\hat{A}}_{p} |$ (\S~\ref{subsec:transition_matrix}).
Consequently, it is more principled to implement the classification and regression heads
as fully connected layers (see Fig.~\ref{fig:algorithmOverview}~\textbf{(f)}).
%In case of lot of consistent (for eg. all cell only move to one step in either direction)
% \SAHA{Suman: not clear:-}
If we observe consistent off-diagonal entries in the transition matrices (e.g. lots of cells moving one step in the same direction),
we could perform pooling as convolution feature map stacking with padding to allow spatial movement.
% in the feature maps.
However, transition matrices are empirically extremely sparse 
(e.g., there are only 25 and 1908 off-diagonal non-zero entries in the transition matrices at $\Delta$ equal to 4 and 20, respectively, on the UCF101-24 dataset).

% Fabio: we should probably expand on this to address the ECCV reviewer criticism
\textbf{Anchor micro-tube sampling.}
% In addition, 
Each transition matrix is converted into a binary one by thresholding, so that the cardinality of the matrix depends not only on the data but also on the transition probability threshold. 
{Our transition matrix based anchor micro-tube sampling scheme is stochastic in nature and emulates Monte Carlo sampling technique (Fig.~\ref{fig:trans_mat} (b)). 
A thresholding on the transition matrix allows us to sample a variable number of anchors rather than a fixed one.}  
We empirically found that a 10\% threshold gives the best results in all of our tests. We discuss the threshold and its effect on performance in \S~\ref{subsec:transition_matrix}. 

The pooling layer (see Fig.~\ref{fig:algorithmOverview}~\textbf{(d)}) is configured to
pool features from a pair of convolutional feature maps
$\{K^{t}_{p}$, $K^{t+\Delta}_{p}\}$ each of shape
$[H'_{p} \times W'_{p} \times D]$.
The pooling is done at cell locations %$loc_1$ and $loc_2$ for feature maps $K^{t}_{p}$ and $K^{t+\Delta}_{p}$
$i$ and $j$,
%respectively, where the cell locations are defined
specified by the estimated (thresholded) transition matrix $\mathbf{\hat{A}}_p$ (\S~\ref{subsec:transition_matrix}).
The pooling kernel has dimension $[3 \times 3 \times D]$.
Pooled features are subsequently stacked
(Fig.~\ref{fig:algorithmOverview}~\textbf{(e)})
to get a single feature representation of a shape
$[2 \times 3 \times 3 \times D]$ per anchor micro-tube.
\iffalse
\textbf{Generality at Test Time.} 
Our approach allows the pooling layer to be \emph{``configurable''}
for a specific set of transition matrices,
while keeping the number of network parameters intact
by simply replacing the convolution heads with linear layers.
Another big advantage of using a transition matrix
% based reconfigurable pooling layer
is that, at test time, we can generate a more generic set of anchor micro-tubes
by setting the transition probabilities to $1$ for those off-diagonal entries
which were earlier $0$ for the training set, or accommodating all anchor micro-tubes transitions relative to each cell.
At test time, in cases in which test $\Delta$ is larger than 1, 
we use linear interpolation to generate the missing detection boxes within a micro-tube.
\fi
\subsection{Classification and regression layers} \label{subsec:conv_2_linear}
%\red{[Note for Gurkirt: please check this section.]}\\
After pooling and stacking, we get $M$ conv features of size
$[2 \times 3 \times 3 \times D]$, for each $M$ anchor micro-tube cell regions
where $M = \sum_{p=1}^{P=6} |\mathbf{\hat{A}}_p|$
is the sum of the cardinalities of the $P$ transition matrices.
We pass these $M$ features to a classification layer
$( (18 \times D)$, $((C+1) \times r) )$,
and a regression layer
$( (18 \times D)$, $((2 \times 4) \times r) )$
(see Fig.~\ref{fig:algorithmOverview}~\textbf{(f)}).
The classification layer outputs $C+1$ class scores
and the regression layer outputs $2 \times 4$ bounding-box coordinates
for $r$ anchor micro-tubes per anchor micro-tube cell region
(see Fig.~\ref{fig:algorithmOverview}~\textbf{(g)}).
The linear classification and regression layers have the
same number of parameters as the convolutional heads in the SSD network~\cite{liu15ssd}.

\subsection{Online action tube generation and temporal trimming}\label{subsub:algo}
The output of the proposed network is a set of detection micro-tubes
and their class confidence scores (see Fig.~\ref{fig:algorithmOverview}~\textbf{(g)}).
We adapt the online action tube generation algorithm
proposed by Singh~\etal~\cite{singh2016online}
to compose these detection micro-tubes into complete action paths (tracklets)
spanning the entire video.
Note that, Singh~\etal~\cite{singh2016online} use their tube generation
algorithm to temporally connect frame-level detection bounding-boxes,
whereas our modified version of the algorithm connects video-level detection
micro-tubes. Similarly to~\cite{singh2016online},
we build action paths incrementally by connecting micro-tubes across time.
as the action paths are extracted,
their temporal trimming %(to solve for temporal detection)
is performed using dynamic programming~\cite{Saha2016,Evangel-2014}.
In Section~\ref{subsec:discussion} we show that
temporal segmentation helps improve
detection performance for datasets
containing highly temporally untrimmed videos
e.g., DALY~\cite{daly2016weinzaepfel},
where on average only $4$\% of the video duration is covered by action instances.
\\
\textbf{Fusion of appearance and flow cues}
We follow a late fusion strategy~\cite{kalogeiton2017action,singh2016online}
to fuse appearance and optical flow cues, performed at test time after all the detections
are extracted from the two streams.
Kalogeiton~\etal~\cite{kalogeiton2017action}
demonstrated that
\textit{mean} fusion works better than both 
\textit{boost} fusion~\cite{Saha2016} and
\textit{union-set} fusion~\cite{singh2016online}.
Thus, in this work we produce all results (cf. Section~\ref{subsec:discussion})
using \textit{mean} fusion~\cite{kalogeiton2017action}.
We report an ablation study of the appearance and flow stream performance
in the supplementary material.
\section{Experiments} \label{subsec:discussion}
We first present datasets, evaluation metrics, fair comparison and 
implementation details used in Section~\ref{subsec:datasets}.
Secondly, we show how TraMNet is able to improve spatial-temporal action localisation in Section~\ref{subsec:st_performance}.
Thirdly, in Section~\ref{subsec:generality}, we discuss how a network learned using transition matrices is able to generalise at test time, when more general anchor-micro-tubes are used to evaluate the network. 
Finally, in Section~\ref{subsec:sparsity}, we quantitatively demonstrate that TraMNet is able to effectively handle sparse annotation as in the DALY dataset, and generalise well on various train and test $\Delta$'s.
\vspace{-0.4cm}
\subsection{Datasets}\label{subsec:datasets}
%We test our approach (\S~\ref{sec:methodology}) on two challenging  benchmarks,
We selected
UCF-101-24~\cite{soomro-2012} to validate the effectiveness of the transition matrix approach, and
DALY~\cite{daly2016weinzaepfel} to evaluate the method on sparse annotations.\\
\textbf{UCF101-24} is a subset of 24 classes from UCF101~\cite{soomro-2012} dataset, which has 101 classes.
Initial 
%It was initially released with 
spatial and temporal annotations provided in 
THUMOS-2013~\cite{idrees2017thumos} were later corrected by Singh \etal~\cite{singh2016online} -- % corrected errors in annotation,
we use this version in all our experiments.
UCF101 videos contain a single action category per video, sometimes
multiple action instances in the same video. 
Each action instance cover on average 70\% of the video duration. % on an average with an average of 1.5 instances in a video.
%The interesting aspect of this dataset according to our method is that 
This dataset is relevant to us as we can show
how the increase in $\Delta$ affects the performance of TraMNet~\cite{saha2017amtnet}, 
and how the transition matrix helps recover from that performance drop.
% \SAHA{Suman: not clear:-}
\textbf{Transformed-UCF101-24} was created by us by padding all images along both the horizontal and the vertical dimension. We set the maximum padding values to
32 and 20 pixels, respectively, as $40\%$ of the average width (80) and height (52) of bounding box annotations.
A uniformly sampled random fraction of 32 pixels is padded on the left edge of the image, the remaining is padded on the right edge of the image. Similar random padding is performed at the top and bottom of each frame. The padding itself is obtained by mirroring the adjacent portion of the image through the edge. % Fabio: maybe we should show an example visually
The same offset is applied to the bounding box annotations.
The \textbf{DALY} dataset was released by Weinzaepfel \etal ~\cite{daly2016weinzaepfel} 
for 10 daily activities and contains 520 videos (200 for test and the rest for training)
with 3.3 million frames. 
Videos in DALY are much longer, and the action duration to video duration ratio is only 4\% compared to UCF101-24's 70\%,
making the temporal labelling of action tubes very challenging.
The most interesting aspect of this dataset is that it is not densely annotated,
as at max 5 frames are annotated per action instance, and % 88\% of action instances have more than one frame annotated remaining
12\% of the action instances only have one annotated frame.
As a result, annotated frames are 2.2 seconds apart on average ($\Delta = 59$). % is 59 on an average i.e.  away from each other on an average.
%These annotated frames contain bounding boxes around actors, and an action label is assigned to each box.
%Along with UCF101-24, we will use the dataset to show how the transition matrix helps in handling sparse annotations.
%\vspace{-0.3cm}
\noindent
\textbf{Note}. THUMOS~\cite{gorban2015thumos} and Activity-Net~\cite{caba2015activitynet} are not suitable for spatiotemporal detection, as they lack bounding box annotation. Annotation at 1fps for AVA~\cite{ava2017gu} was released in week 1 of March 2018 (to the best of our knowledge). Also, AVA's bounding boxes are not linked in time, preventing a fair evaluation of our approach there.
\\
\textbf{\textit{Evaluation metric.}} \label{para:eval_metric}
We evaluate TraMNet using video-mAP \cite{peng2016eccv,yu2015fast,singh2016online,kalogeiton2017action,saha2017amtnet}.
As a standard practice~\cite{singh2016online}, 
we use \emph{``average detection performance''} (avg-mAP) to compare TraMNet's performance 
with the state-of-the-art.
To obtain the latter, we first compute the video-mAPs at higher IoU thresholds ($\delta$) 
ranging $[0.5:0.05:0.95]$, and then take the average of these video-mAPs.
% Similar to ~\cite{singh2016online}, we pick standard comparison metric as average-mAP (mean average precision) for evaluation threshold ($\delta$) ranging from 0.5 to 0.95 at the gap of 0.05 on UCF101-24 dataset. 
On the DALY dataset, we also evaluate at various thresholds in both an untrimmed and a trimmed setting. 
The latter is achieved by trimming the action paths generated by the boundaries of the ground truth~\cite{daly2016weinzaepfel}. 
We further report the video classification accuracy using the predicted tubes as in~\cite{singh2016online}, 
in which videos are assigned the label of the highest scoring tube.
One can improve classification on DALY by 
taking into consideration of other tube scores.
Nevertheless, in our tests we adopt the existing protocol.
\\
For \textbf{\textit{fair comparison}} \label{para:fair_compar}
we re-implemented the methods of our competitors~\cite{Saha2016,kalogeiton2017action,singh2016online}
with SSD as the base network. As in our TraMNet network, 
we also replaced SSD's convolutional heads with new linear layers. 
The same tube generation~\cite{singh2016online} and data augmentation~\cite{liu15ssd} methods were adopted,
and the same hyperparameters were used for training all the networks, including TraMNet.
The only difference is that the anchor micro-tubes used in ~\cite{Saha2016,kalogeiton2017action} were cuboidal, 
whereas TraMNet's anchor micro-tubes are generated using transition matrices.
We refer to these approaches as SSD-L (SSD-linear-heads)~\cite{singh2016online}, AMTnet-L (AMTnet-linear-heads) 
~\cite{saha2017amtnet} and as ACT-L (ACT-detector-linear-heads)~\cite{kalogeiton2017action}.
\\
\noindent
\textbf{Network training and implementation details.}
We used the established training settings for all the above methods.
While training on the UCF101-24 dataset,
we used a batch size of $16$ and an initial learning rate of $0.0005$, with the
learning rate dropping after $100K$ iterations for the appearance stream and $140K$ for the flow stream.
Whereas the appearance stream is only trained for $180K$ iterations,
the flow stream is trained for $200K$ iterations.
In all cases, the input image size was $3\times 300\times 300$ for the appearance stream, while a
stack of five optical flow images~\cite{Brox-2004} ($15\times 300\times 300$) was used for flow.
Each network was trained on $2$ 1080Ti GPUs.
More details about parameters and training are given in the supplementary material.
%-----------------------------------------------------------------------------------------------------------------
% UCF-101 results TABLE
%-----------------------------------------------------------------------------------------------------------------
\begin{table}[t]
  %\vskip -3mm
  \centering
  \setlength{\tabcolsep}{3.5pt}
  \caption{Action localisation results on untrimmed videos from UCF101-24 split1.
  The table is divided into 4 parts. The first part lists approaches which have single frames as input;
  the second part approaches which take multiple frames as input;
  the third part contemplates the re-implemented versions of approaches in the second group;
  lastly, we report our TraMNet's performance.}
  \vspace{-0.2cm}
  {\footnotesize
  \scalebox{0.9}{
  \begin{tabular}{lccccccc}
  \toprule
  Methods & Train $\Delta$ & Test $\Delta$	&$\delta$ = 0.2   &$\delta$ = 0.5 & $\delta$ = 0.75 & $\delta$ = .5:.95 & Acc \% \\ \midrule
  %Yu \etal~\cite{yu2015fast} &NA &NA & 26.5  & --   & -- & -- & --\\
  %Weinzaepfel ~\cite{Weinzaepfel-2015}  & NA &NA & 46.8  & --   & -- & -- & --\\
  T-CNN~\cite{hou2017tube} & NA &NA & 47.1  & --   & -- & -- & --\\
  MR-TS~\cite{peng2016eccv}	 & NA &NA & 73.5  & 32.1 & 02.7 & 07.3 & --\\
  Saha~\etal~\cite{Saha2016}    & NA &NA & 66.6  & 36.4 & 07.9 & 14.4 & --\\
  SSD~\cite{singh2016online}    & NA & NA & 73.2 & 46.3 & 15.0 & 20.4 & --\\\midrule
  AMTnet~\cite{saha2017amtnet} rgb-only  & 1,2,3 & 1 & 63.0  & 33.1 & 00.5 & 10.7 & --\\
  ACT~\cite{kalogeiton2017action}   & 1 & 1 & 76.2  & 49.2 & 19.7 & 23.4 & --\\
  Gu \etal~\cite{ava2017gu} (\cite{peng2016eccv} + \cite{carreira2017quo})   & NA & NA & --  & \textbf{59.9} & -- & -- & --\\
  \midrule
  SSD-L with-trimming  & NA & NA & 76.2 & 45.5 & 16.4 & 20.6  & 92.0 \\
  SSD-L     & NA & NA & 76.8 & 48.2 & 17.0 & 21.7  & 92.1 \\
  ACT-L     & 1 & 1 & 77.9 & 50.8 & 19.8 & \textbf{23.9}  & 91.4 \\
  AMTnet-L  & 1 & 1 & \textbf{79.4} & \textbf{51.2} & 19.0 & 23.4  & \textbf{92.9} \\
  %AMTnet-L  & 0 & 4 & 76.9 & 48.5 & 16.5 & 21.5  & 92.1 \\
  AMTnet-L  & 5 & 5 & 77.5 & 49.5 & 17.3 & 22.5  & 91.6 \\
  %AMTnet-L  & 9 & 4 & 76.1 & 48.9 & 17.7 & 22.2  & 91.2 \\
  AMTnet-L  & 21 & 5 & 76.2 & 47.6 & 16.5 & 21.6  & 90.0 \\\midrule
  TraMNet (ours)& 1 & 1 & 79.0 & 50.9 & \textbf{20.1} & \textbf{23.9}  & 92.4 \\
  %TraMNet (ours)& 0 & 4 & 76.9 & 47.0 & 16.9 & 21.6  & 91.5 \\
  TraMNet (ours)& 5 & 5 & 77.6 & 49.7 & 18.4 & 22.8  & 91.3 \\
  %TraMNet (ours)& 9 & 4 & 76.5 & 50.1 & 16.9 & 22.6  & 92.6 \\
  TraMNet (ours)& 21 & 5 & 75.2 & 47.8 & 17.4 & 22.3  & 90.7 \\ \bottomrule
  \end{tabular}
  }
  }
  %\vspace*{-\baselineskip}
  \label{table:ucf101_results} \vspace{-6mm}
\end{table}
%%------------------------------------------------------------
%%%----------------------------------------------------------
%%------------------------------------------------------------
%%%----------------------------------------------------------
\subsection{Action localisation performance}\label{subsec:st_performance}

Table~\ref{table:ucf101_results} shows the resulting performance on UCF101-24 at multiple train and test $\Delta$s for TraMNet versus other competitors~\cite{Saha2016,kalogeiton2017action,singh2016online,peng2016eccv,hou2017tube}.
Note that Gu~\etal~\cite{ava2017gu} build upon MS-TS~\cite{peng2016eccv} by adding a strong I3D~\cite{carreira2017quo} base network, making it unfair to compare \cite{ava2017gu} to SSD-L, AMTnet-L, ACT-L and TraMNet, which all use VGG as a base network.
\\
ACT is a dense network (processin $6$ consecutive frames), which
shows the best performance at high overlap (an avg-mAP of 23.9\%).
AMTnet-L is slightly inferior ($23.4$\%), 
most likely due to it learning representations from pairs of 
consecutive frames only at its best training and test settings ($\Delta = 1$).
TraMNet is able to match ACT-L's performance at high overlap ($23.9$\%), while being comparatively more efficient.

% Fabio: is this all we have on TransformedUCF? A bit disappointing, I feel it should be given more space
The evaluation of AMTNet-L on Transformed-UCF101-24 (\S ~\ref{subsec:datasets}) shows an avg-mAP of $19.3\%$ using the appearance stream only, whereas
TraMNet records an avg-mAP of $20.5\%$, a gain of $1.2\%$ that can be attributed to its estimating grid location transition probabilities. 
It shows that TraMNet is more suited to action instances involving substantial shifts from one frame to the next. 
A similar phenomenon can be observed on the standard UCF101-24 when the train or test $\Delta$ is greater than $1$ in Table~\ref{table:ucf101_results}.

We cross-validated different transition probability thresholds on transition matrices. Thresholds of $2\%$, $5\%$, $10\%$, $15\%$ and $20\%$ yielded an avg-mAP of $21.6\%$, $22.0\%$, $22.4\%$, $21.9\%$ and $21.2\%$, respectively, on the appearance stream. Given such evidence, we concluded that a $10\%$ transition probability threshold was to be adopted throughout all our experiments.

%Fabio: I think this paragraph is quite unclear, and uses a sloppy language
%\subsection{Generality of Anchors Micro-Tubes at Test Time}\label{subsec:generality}
\subsection{Location invariance at test time}\label{subsec:generality}

{Anchor micro-tubes are sampled based on the transition probabilities 
from specific cells (at frame $f_t$) to other specific 
cells (at frame $f_{t+\Delta}$) (\S~\ref{subsec:transition_matrix}) based on the training data.
However, as at test time action instances of a same class may appear 
in other regions of the image plane than those observed at training time,
it is desirable to generate additional anchor micro-tubes proposals than those produced by the learnt transition matrices.
Such \emph{location invariance} property can be achieved at test time by augmenting 
the binary transition matrix (\S~\ref{subsec:reconfig_pooling_layer}) 
with likely transitions from other grid locations.}

% subsec:transition_matrix
% $\mathbf{\hat{A}}[i,j] = 0$.
{Each row/column of the transition matrix $\mathbf{\hat{A}}$ (\S~\ref{subsec:transition_matrix}) 
corresponds to a cell location in the grid.
One augmentation technique is to set all the diagonal entries to 1 (i.e., $\mathbf{\hat{A}}[i,j] = 1$, where $i==j$). This amounts to generating anchor cuboids 
which may have been missing at training time (cfr. Fig.~\ref{fig:trans_mat} (a)).
The network can then be evaluated using this new set of anchor micro-tubes 
by configuring the pooling layer (\S~\ref{subsec:reconfig_pooling_layer})) accordingly.
% as the pooling layer (\S~\ref{subsec:reconfig_pooling_layer})) can be configured accordingly. 
When doing so, however, we observed only a very minor difference in avg-mAP at the second decimal point 
for TraMNet with test $\Delta=1$.
Similarly, we also evaluated TraMNet by incorporating the transitions from each cell to its 8 \emph{neighbouring} cells (also at test time), but observed no significant change in avg-mAP.}

A third approach,
%\textbf{Relative Anchors:} 
{given a pyramid level $p$, and the initial binary transition matrix for that level, 
% (in each row, the difference in the position from diagonal entry to the column with entry equal to 1 i.e. possible transition) 
consists of computing the relative transition offsets for all grid cells (offset $= i-j \ \forall i,j$ where  $\mathbf{\hat{A}}[i,j] = 1$).
All such transition offsets correspond to different spatial translation patterns (of action instances) present in the dataset at different locations in the given video.
Augmenting all the rows with these spatial translation patterns, by taking each diagonal entry in the transition matrix as reference point, yields a more dense transition matrix whose anchor micro-tubes are translation invariant, i.e., spatial location invariant.
However, after training TraMNet at train $\Delta = 1$ we observed that the final avg-mAP at test $\Delta = 1$ was $22.6\%$ as compared to $23.9\%$ when using the original (sparse) transition matrix. 
As in the experiments (i.e., added diagonal and neighbour transitions) explained above, we evaluated the network that was trained on the original transition matrices at train $\Delta=1$  by using the transition matrix generated via relative offsets, observing an avg-mAP consistent (i.e., $23.9\%$) with the original results. 

This shows that the system should be trained using the original transition matrices learned from the data, whereas more anchor micro-tube proposals can be assessed at test time without loss of generality.}
%SAVING SPACE:::--- Hence, we can train our network efficiently and evaluate with more anchors as evaluation is much faster than training.
It also shows that UCF101-24 is not sufficiently realistic a dataset from the point of view of translation invariance, which is why we conducted tests on Transformed-UCF101-24 (\S ~\ref{subsec:datasets}) to highlight this issue. % Fabio: but then we are quite glossing over the new datasets and its results

\begin{table}[t]
  %\vskip -3mm
  \centering
  \caption{Action localisation results (video-mAP) on the DALY dataset. SSD-L without trimming refers to when action paths are not trimmed and the network is SSD.}
  \vspace{-0.2cm}
  {\footnotesize
  \scalebox{0.9}{
  \begin{tabular}{lc|ccc|cccc}
  \toprule
  \multicolumn{2}{c}{} &\multicolumn{3}{|c|}{Untrimmed Videos} & \multicolumn{4}{c}{Trimmed Videos} \\
  Methods & Test $\Delta$	&  $\delta$=0.2   & $\delta$=0.5 & Acc\% & $\delta$=0.5 & $\delta$=.5:.95 & Acc\% & CleaningFloor \\\midrule
  weinzaepfel \etal ~\cite{daly2016weinzaepfel} & NA & 13.9 & -- & -- & 63.9   & -- & -- & --\\
  SSD-L without-trimming & NA & 06.1 & 01.1 & 61.5 & \multicolumn{4}{c}{}\\
  SSD-L & NA & \textbf{14.6} & \textbf{05.7} & 58.5 & 63.9 & 38.2  & 75.5 & 80.2\\
  AMTnet-L & 3 & 12.1 & 04.3 & 62.0  & 63.7 & 39.3 & 76.5 & 83.4\\
  TraMNet (ours) & 3 & 13.4 & 04.6 & \textbf{67.0} & \textbf{64.2}  & \textbf{41.4}  & \textbf{78.5} & \textbf{86.6}\\ \bottomrule
  \end{tabular}
  }
  }
  %\vspace*{-\baselineskip}
  \label{table:daly_results} \vspace{-6mm}
\end{table}

\subsection{Handling sparse annotations}\label{subsec:sparsity}

Table~\ref{table:daly_results} shows the results on the DALY dataset.
We can see that TraMNet significantly improves on SSD-L and AMTnet-L in the trimmed video setting, 
with an avg. video-mAP of $41.4$\%.
TraMNet reaches top classification accuracy in both the trimmed and the untrimmed cases.
As we would expect, TraMNet improves the temporal linking via better micro-tubes and classification, 
as clearly indicated in the trimmed videos setting.
Nevertheless, SSD-L is the best when it comes to temporal trimming.
We think this is because each micro-tube in 
our case is 4 frames long as the test $\Delta$ is equal to 3, %Fabio: it says 3 in the table
and each micro-tube only has one score vector rather than 4 score vectors for each frame,
which might smooth temporal segmentation aspect. %Fabio: I don't quite get what you mean
\\
DALY allows us to show how TraMNet is able to handle 
sparse annotations better than AMTNet-L, which uses anchor cuboids, %In DALY, at training time $\Delta$ is arbitrary as only max 5 frames are annotated for any given action instance. As a result, 
%TraMNet performs better than AMTnet-L in both settings, clearly 
strengthening the argument that learning transition matrices helps generate better micro-tubes.

\begin{figure}[t]
    \vspace{-0.2cm}
    \centering
    \includegraphics[width=0.96\textwidth]{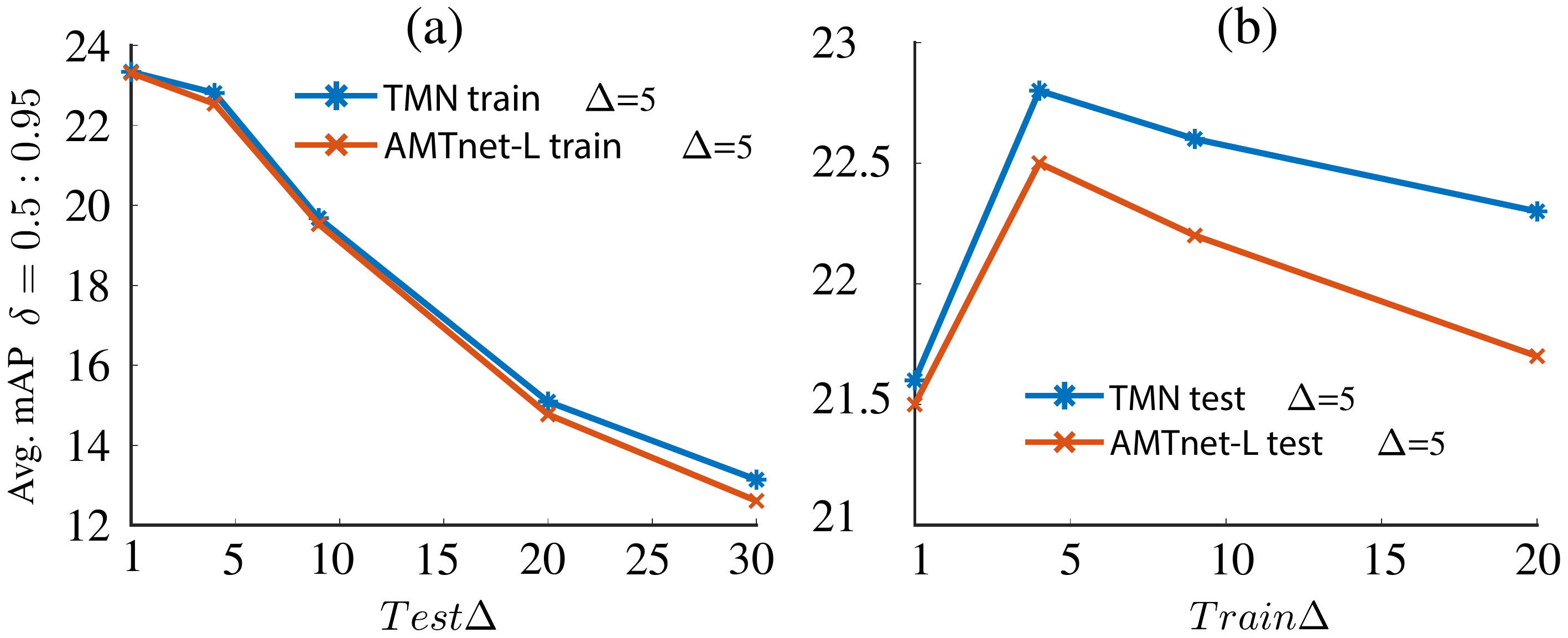}
    \vspace{-0.4cm}
    \caption{Avg mAP ($\delta=0.5:0.95$) performance of TraMNet vs the competitor AMTnet-L,
    (a) when tested at  constant $\Delta$ equal to 5 and trained on increasing $\Delta$ from 1 to 20,
    (b) when tested at increasing $\Delta$ from 1 to 30 and trained at constant $\Delta$ equal to 5.}
    \label{fig:deltas}
    \vspace{-0.5cm}
\end{figure}

TramNet's performance on `CleaningFloor' at $\delta$ equal to 0.5 in 
the trimmed case highlights the effectiveness of general anchor micro-tubes for \textbf{dynamic classes}. 
`CleaningFloor' is one of DALY's classes in which the actor moves spatially while the camera is mostly static. 
To further strengthen the argument, 
we picked classes showing fast spatial movements across frames in the UCF101-24 dataset
and observed the class-wise average-precision (AP) at $\delta$ equal to $0.2$.
For `BasketballDunk', `Skiing' and `VolleyballSpiking' 
TraMNet performs significantly better than both AMTnet-L and ACT-L; e.g. 
on `Skiing', the performance of TraMNet, 
AMTNet-L and ACT-L is $85.2$, $82.4$ and $81.1$, respectively.
%We picked a lower $\delta=0.2$ to limit the effect of temporal labelling, to be able to
%verify the quality of the micro-tubes for all methods.
More class-wise results are discussed in the supplementary material.
\\
\textbf{Training and testing at multiple $\Delta$'s}\label{subsec:deltas}
To test whether TraMNet can handle sparse annotation 
we introduced an artificial gap ($\Delta$) in UCF101's training examples, 
while testing on frames that are far away (e.g. $\Delta = 30$).
We can observe in Figure \ref{fig:deltas}(a) 
that performance is preserved when increasing the training $\Delta$ 
while keeping the test $\Delta$ small (e.g. equal to 5, as shown in plot (a)).
One could think of increasing $\Delta$ at test time to improve run-time efficiency: 
we can observe from Figure \ref{fig:deltas}(b) that performance drops linearly as speed linearly increases.
In both cases TraMNet consistently outperforms AMTNet.
When $\Delta$ is large TraMNet's improvement is large as well. 

\noindent
\textbf{Temporal labelling} is performed using the labelling formulation presented in \cite{singh2016online}.
Actually, temporal labelling hurts the performance on UCF101-24, 
as shown in Table \ref{table:ucf101_results} where `SSD-L-with-trimming' 
uses \cite{singh2016online}'s temporal segmenter, whereas `SSD-L' and 
the other methods below that do not.
In contrast, on DALY the results are quite the opposite:
the same temporal labelling framework improves the performance from $6.1\%$ to $14.9\%$ at $\delta = 0.2$.
We think that these (superficially) contradictory results relate to 
the fact that action instances cover on average a very different 
fraction (70\% versus 4\%) of the video duration in UCF101-24 and DALY, respectively.
\\
% \subsection{Test time detection speed}
% \label{sec:detection-speed}
% Singh \etal ~\cite{singh2016online} showcase their method's online and real-time capabilities.
% Here we use their online tube generation method in order for our TraMNet to inherit those properties.
% The only question mark is TraMNet's forward pass speed.
\textbf{Detection speed:} We measured the average time taken for a forward pass for 
a batch size of 1 as compared to 8 by ~\cite{singh2016online}. A single-stream forward 
pass takes 29.8 milliseconds (i.e. 33fps) on a single 1080Ti GPU. 
One can improve speed even further by evaluating TraMNet with $\Delta$ equal 
to 2 or 4,  obtaining a 2$\times$ or 4$\times$ speed improvement 
while paying very little in terms of performance, as shown in Figure \ref{fig:deltas}(b).

\section{Conclusions} \label{sec:conclusions}
We presented a TraMNet deep learning framework for action detection in videos which,
unlike 
previous state-of-the-art methods~\cite{saha2017amtnet,kalogeiton2017action,hou2017tube} 
which generate action cuboid proposals,
%by extending the 2D anchor boxes to their 3D counterparts (i.e. anchor cuboids).
%This way of generating proposals is suitable for object detection
%in still images or for ``static'' actions where the actor 
%does do not change location in time (e.g. ``clapping'' action).
%However in 
can cope with real-world videos containing ``dynamic'' actions whose location significantly changes over time. This is done by learning a transition probability matrix for each feature pyramid layer from the training data in a hidden Markov model formulation, leading to an original configurable layer architecture.
%majority of the actions are ``dynamic'' and thus a cuboidal shape proposal is not an optimal solution for this task.
Furthermore, unlike its competitors %the state-of-the-art methods~
\cite{kalogeiton2017action,hou2017tube}, which
require dense frame-level bounding box annotation,
%for $k$ ($6$ or $8$) consecutive frame which is expensive for large-scale datasets such as AVA or DALY dataset.
TraMNet builds on the network architecture of~\cite{saha2017amtnet}
in which action representations are learnt from pairs of frames rather than chunks of consecutive frames, thus eliminating the need for dense annotation.
An extensive experimental analysis supports TraMNet's action detection capabilities, especially
%specifically for long real-world video
under dynamic actions and sparse annotations. 
%%%%%%%%%%%%%%%%%%%%%%%%%%%%%%%%%%%%%%%%%%%%%%
% \clearpage
\bibliographystyle{splncs03}
\bibliography{bib}

\begin{thebibliography}{10}
\providecommand{\url}[1]{\texttt{#1}}
\providecommand{\urlprefix}{URL }

\bibitem{Brox-2004}
Brox, T., Bruhn, A., Papenberg, N., Weickert, J.: High accuracy optical flow
  estimation based on a theory for warping (2004)

\bibitem{caba2015activitynet}
Caba~Heilbron, F., Escorcia, V., Ghanem, B., Carlos~Niebles, J.: Activitynet: A
  large-scale video benchmark for human activity understanding. In: {IEEE} Int.
  Conf. on Computer Vision and Pattern Recognition. pp. 961--970 (2015)

\bibitem{carreira2017quo}
Carreira, J., Zisserman, A.: Quo vadis, action recognition? a new model and the
  kinetics dataset. In: 2017 IEEE Conference on Computer Vision and Pattern
  Recognition (CVPR). pp. 4724--4733. IEEE (2017)

\bibitem{elliott2008hidden}
Elliott, R.J., Aggoun, L., Moore, J.B.: Hidden Markov models: estimation and
  control, vol.~29. Springer Science \& Business Media (2008)

\bibitem{Evangel-2014}
Evangelidis, G., Singh, G., Horaud, R.: Continuous gesture recognition from
  articulated poses. In: ECCV Workshops (2014)

\bibitem{vanGemert2015apt}
van Gemert, J.C., Jain, M., Gati, E., Snoek, C.G.: {APT}: Action localization
  proposals from dense trajectories. In: BMVC. vol.~2, p.~4 (2015)

\bibitem{girshick-2014}
Girshick, R., Donahue, J., Darrel, T., Malik, J.: Rich feature hierarchies for
  accurate object detection and semantic segmentation. In: {IEEE} Int. Conf. on
  Computer Vision and Pattern Recognition (2014)

\bibitem{Georgia-2015a}
Gkioxari, G., Malik, J.: Finding action tubes. In: {IEEE} Int. Conf. on
  Computer Vision and Pattern Recognition (2015)

\bibitem{gorban2015thumos}
Gorban, A., Idrees, H., Jiang, Y., Zamir, A.R., Laptev, I., Shah, M.,
  Sukthankar, R.: Thumos challenge: Action recognition with a large number of
  classes (2015)

\bibitem{ava2017gu}
Gu, C., Sun, C., Vijayanarasimhan, S., Pantofaru, C., Ross, D.A., Toderici, G.,
  Li, Y., Ricco, S., Sukthankar, R., Schmid, C., et~al.: Ava: A video dataset
  of spatio-temporally localized atomic visual actions. arXiv preprint
  arXiv:1705.08421  (2017)

\bibitem{he2016deep}
He, K., Zhang, X., Ren, S., Sun, J.: Deep residual learning for image
  recognition. In: Proceedings of the IEEE conference on computer vision and
  pattern recognition. pp. 770--778 (2016)

\bibitem{hou2017tube}
Hou, R., Chen, C., Shah, M.: Tube convolutional neural network (t-cnn) for
  action detection in videos. In: {IEEE} Int. Conf. on Computer Vision (2017)

\bibitem{idrees2017thumos}
Idrees, H., Zamir, A.R., Jiang, Y.G., Gorban, A., Laptev, I., Sukthankar, R.,
  Shah, M.: The thumos challenge on action recognition for videos “in the
  wild”. Computer Vision and Image Understanding  155,  1--23 (2017)

\bibitem{jain2014tublet}
Jain, M., Van~Gemert, J., J{\'e}gou, H., Bouthemy, P., Snoek, C.G.: Action
  localization with tubelets from motion. In: Computer Vision and Pattern
  Recognition (CVPR), 2014 IEEE Conference on. pp. 740--747. IEEE (2014)

\bibitem{kalogeiton2017action}
Kalogeiton, V., Weinzaepfel, P., Ferrari, V., Schmid, C.: Action tubelet
  detector for spatio-temporal action localization. In: {IEEE} Int. Conf. on
  Computer Vision (2017)

\bibitem{kay2017kinetics}
Kay, W., Carreira, J., Simonyan, K., Zhang, B., Hillier, C., Vijayanarasimhan,
  S., Viola, F., Green, T., Back, T., Natsev, P., et~al.: The kinetics human
  action video dataset. arXiv preprint arXiv:1705.06950  (2017)

\bibitem{lin2017focal}
Lin, T.Y., Goyal, P., Girshick, R., He, K., Doll{\'a}r, P.: Focal loss for
  dense object detection. arXiv preprint arXiv:1708.02002  (2017)

\bibitem{liu15ssd}
Liu, W., Anguelov, D., Erhan, D., Szegedy, C., Reed, S., Fu, C.Y., Berg, A.C.:
  {SSD}: Single shot multibox detector. arXiv preprint arXiv:1512.02325  (2015)

\bibitem{oneata2014efficient}
Oneata, D., Verbeek, J., Schmid, C.: Efficient action localization with
  approximately normalized fisher vectors. In: Proceedings of the IEEE
  Conference on Computer Vision and Pattern Recognition. pp. 2545--2552 (2014)

\bibitem{peng2016eccv}
Peng, X., Schmid, C.: {Multi-region two-stream R-CNN for action detection}. In:
  {ECCV 2016 - European Conference on Computer Vision}. Amsterdam, Netherlands
  (Oct 2016), \url{https://hal.inria.fr/hal-01349107}

\bibitem{redmon2016yolo9000}
Redmon, J., Farhadi, A.: Yolo9000: Better, faster, stronger. arXiv preprint
  arXiv:1612.08242  (2016)

\bibitem{ren2015faster}
Ren, S., He, K., Girshick, R., Sun, J.: Faster {R-CNN}: Towards real-time
  object detection with region proposal networks. In: Advances in Neural
  Information Processing Systems. pp. 91--99 (2015)

\bibitem{saha2017amtnet}
Saha, S., Singh, G., Cuzzolin, F.: Amtnet: Action-micro-tube regression by
  end-to-end trainable deep architecture. In: {IEEE} Int. Conf. on Computer
  Vision (2017)

\bibitem{Saha2016}
Saha, S., Singh, G., Sapienza, M., Torr, P.H.S., Cuzzolin, F.: Deep learning
  for detecting multiple space-time action tubes in videos. In: British Machine
  Vision Conference (2016)

\bibitem{sapienza2014}
Sapienza, M., Cuzzolin, F., Torr, P.H.: Learning discriminative space-time
  action parts from weakly labelled videos. Int. Journal of Computer Vision
  (2014)

\bibitem{singh2016online}
Singh, G., Saha, S., Sapienza, M., Torr, P., Cuzzolin, F.: Online real-time
  multiple spatiotemporal action localisation and prediction. In: {IEEE} Int.
  Conf. on Computer Vision (2017)

\bibitem{soomro-2012}
Soomro, K., Zamir, A.R., Shah, M.: {UCF101}: A dataset of 101 human action
  classes from videos in the wild. Tech. rep., CRCV-TR-12-01 (2012)

\bibitem{Sultani2016}
Sultani, W., Shah, M.: What if we do not have multiple videos of the same
  action? - video action localization using web images. In: {IEEE} Int. Conf.
  on Computer Vision and Pattern Recognition (2016)

\bibitem{wang-2011}
Wang, H., Kl{\"a}ser, A., Schmid, C., Liu, C.: {Action Recognition by Dense
  Trajectories}. In: {IEEE} Int. Conf. on Computer Vision and Pattern
  Recognition (2011)

\bibitem{Weinzaepfel-2015}
Weinzaepfel, P., Harchaoui, Z., Schmid, C.: {Learning to track for
  spatio-temporal action localization}. In: {IEEE} Int. Conf. on Computer
  Vision and Pattern Recognition (June 2015)

\bibitem{daly2016weinzaepfel}
Weinzaepfel, P., Martin, X., Schmid, C.: Human action localization with sparse
  spatial supervision. arXiv preprint arXiv:1605.05197  (2016)

\bibitem{weinzaepfel2016towards}
Weinzaepfel, P., Martin, X., Schmid, C.: Towards weakly-supervised action
  localization. arXiv preprint arXiv:1605.05197  (2016)

\bibitem{yang2017spatio}
Yang, Z., Gao, J., Nevatia, R.: Spatio-temporal action detection with cascade
  proposal and location anticipation. In: BMVC (2017)

\bibitem{yu2015fast}
Yu, G., Yuan, J.: Fast action proposals for human action detection and search.
  In: Proceedings of the IEEE Conference on Computer Vision and Pattern
  Recognition. pp. 1302--1311 (2015)

\bibitem{zolfaghari2017chained}
Zolfaghari, M., Oliveira, G.L., Sedaghat, N., Brox, T.: Chained multi-stream
  networks exploiting pose, motion, and appearance for action classification
  and detection. In: {IEEE} Int. Conf. on Computer Vision. pp. 2923--2932. IEEE
  (2017)

\end{thebibliography}
\end{document}